\documentclass{moriond}

\def\be{\begin{equation}}
\def\ee{\end{equation}}
\def\bea{\begin{eqnarray}}
\def\eea{\end{eqnarray}}

\begin{document}
\vspace*{4cm}
\title{FLAVOUR FROM THE PLANCK TO THE ELECTROWEAK SCALE
\footnote{Contribution to the 2019 EW session of the 54th Rencontres de Moriond.}}

\author{ S.F. KING }

\address{Department of Physics and Astronomy, University of Southampton,\\
Southampton, SO17 1BJ, U.K.}

\maketitle\abstracts{
We discuss a theory of flavour in which Higgs Yukawa couplings are related to those of the new scalar triplet 
leptoquark and/or $Z'$
responsible for $R_{K^{(*)}}$, with all couplings arising effectively from mixing with a vector-like fourth family,
whose mass may be anywhere from the Planck scale to the electroweak scale for the leptoquarks explanation, but 
is pinned down to the TeV scale if the $Z'$ exchange plays a role. However, in this particular model, only leptoquark exchange can contribute significantly to $R_{K^{(*)}}$, since $Z'$ exchange is too constrained from $B_s$ mixing and
$\tau \rightarrow 3 \mu$, although other Higgs Yukawa matrix structures may allow it.}

\section{Introduction}
The Standard Model (SM) has almost thirty parameters, most of them arising from the unspecified Higgs Yukawa couplings. This motivates theories of flavour beyond the SM. 
Here we shall consider one simple example 
where the usual Higgs Yukawa couplings with the three families of chiral fermions are forbidden
by some symmetry, which is broken by some new scalar field $\langle \phi \rangle$, allowing 
the Higgs Yukawa couplings to arise effectively from mixing with a vector-like fourth 
family~\cite{Ferretti:2006df}. 
In principle the mass of the vector-like fourth family $M_4$ (i.e. the flavour scale in this example)
may be anywhere
from the Planck scale to the electroweak scale, providing that the ratios $\langle \phi \rangle /M_4$ which govern the Yukawa couplings are held fixed.

The LHCb Collaboration~\cite{moriond} 
and other experiments have reported a number of anomalies in $B\rightarrow K^{(*)}l^+l^-$ 
decays such as the $R_K$ and $R_{K^*}$ ratios of $\mu^+ \mu^-$ to $e^+ e^-$ final states, 
which are observed to be about $80\%$ of their expected values with a $2.5 \sigma$ deviation from the SM.
Such anomalies may be accounted for by a new physics operator of the form~\cite{moriond}  
$\bar b_L\gamma^{\mu} s_L \, \bar \mu_L \gamma_{\mu} \mu_L$,
with a coefficient $\Lambda^{-2}$ where $\Lambda \sim 30$ TeV.
This hints that there may be new physics arising from the non-universal couplings of 
leptoquarks and/or $Z'$ in order to generate such an operator.
However the introduction of these new particles increases the parameter count still further,
and only serves to make the favour problem of the SM worse.

Motivated by such considerations, it is interesting to speculate that the above empirical hint of 
flavour non-universality is linked
to a possible theory of flavour. In this talk we consider an example of this based on the vector-like fourth family
discussed above.
To achieve the desired link, one may introduce leptoquarks and/or $Z'$ into the above theory of flavour 
in such a way that the effective Higgs Yukawa couplings and the effective leptoquark and/or $Z'$ couplings are 
generated at the same time, from mixing with the vector-like fourth family. In such a model, the couplings of 
leptoquarks and/or $Z'$ may be related to Yukawa couplings, leading to a very predictive framework.

\section{Model of flavour and $R_{K^{(^*)}}$ with leptoquarks and $Z'$ }
Consider the model in Table~\ref{model} 
with a vector-like fourth family of fermions of mass $M_4$~\cite{Ferretti:2006df}.
The model also involves a gauged $U(1)'$,
which is broken by a singlet $\phi$ leading to a massive $Z'$ with non-universal couplings~\cite{King:2017anf,Falkowski:2018dsl,King:2018fcg}. We have also included 
a scalar leptoquark triplet $S_3$ of mass $M_{S_3}$~\cite{deMedeirosVarzielas:2018bcy,deMedeirosVarzielas:2019okf}.
The model in Table~\ref{model}, defined in these proceedings for the first time, may be regarded as 
an amalgamation of the $Z'$ model~\cite{King:2018fcg}
and the leptoquark model~\cite{deMedeirosVarzielas:2019okf}, where both models previously included also a
vector-like fourth family of fermions. The idea is that the usual three chiral families of quarks and leptons do not have
renormalisable couplings to Higgs or leptoquarks or $Z'$ (since they are neutral under $U(1)'$). However, as we shall see,
such couplings are generated via mixing with the vector-like fourth family, thereby relating all these couplings to each other.

\begin{table}[ht]
\begin{minipage}[b]{0.56\linewidth}
{\tiny
\centering
\begin{tabular}{| l c c c c |}
\hline
Field & $SU(3)_c$ & $SU(2)_L$ & $U(1)_Y$ &$U(1)'$\\ 
\hline \hline
$Q_{i}$ 		 & ${\bf 3}$ & ${\bf 2}$ & $1/6$ & $0$ \\
$u^c_{i}$ 		 & ${\overline{\bf 3}}$ & ${\bf 1}$ & $-2/3$ & $0$\\
$d^c_{i}$ 		 & ${\overline{\bf 3}}$ & ${\bf 1}$ & $1/3$ & $0$\\
$L_{i}$ 		 & ${\bf 1}$ & ${\bf 2}$ & $-1/2$ & $0$\\
$e^c_{i}$ 		 & ${\bf 1}$ & ${\bf 1}$ & $1$ & $0$\\
$\nu^c_{i}$         & ${\bf 1}$ & ${\bf 1}$ & $0$ & $0$\\
\hline
\hline
$Q_{4}$ 		 & ${\bf 3}$ & ${\bf 2}$ & $1/6$ & $1$\\
$u^c_{4}$ 		 & ${\overline{\bf 3}}$ & ${\bf 1}$ & $-2/3$ & $1$\\
$d^c_{4}$ 		 & ${\overline{\bf 3}}$ & ${\bf 1}$ & $1/3$ & $1$\\
$L_{4}$ 		 & ${\bf 1}$ & ${\bf 2}$ & $-1/2$ & $1$\\
$e^c_{4}$ 		 & ${\bf 1}$ & ${\bf 1}$ & $1$ & $1$\\
$\nu^c_{4}$         & ${\bf 1}$ & ${\bf 1}$ & $0$ & $1$\\
\hline
\hline
$\overline{Q_{4}}$ 		 & $\overline{{\bf 3}}$ & $\overline{{\bf 2}}$ & $-1/6$ & $-1$\\
$\overline{u^c_{4}}$ 		 & ${{\bf 3}}$ & ${\bf 1}$ & $2/3$ & $-1$\\
$\overline{d^c_{4}}$ 		 & ${{\bf 3}}$ & ${\bf 1}$ & $-1/3$ & $-1$\\
$\overline{L_{4}}$ 		 & ${\bf 1}$ & $\overline{{\bf 2}}$ & $1/2$ & $-1$\\
$\overline{e^c_{4}}$ 		 & ${\bf 1}$ & ${\bf 1}$ & $-1$ & $-1$\\
$\overline{\nu^c_{4}}$         & ${\bf 1}$ & ${\bf 1}$ & $0$ & $-1$\\
\hline
\hline
$H_u$ & ${\bf 1}$ & ${\bf 2}$ & $1/2$ &$-1$ \\
$H_d$ & ${\bf 1}$ & ${\bf 2}$ & $-1/2$ & $-1$\\
\hline
\hline
$\phi$ & ${\bf 1}$ & ${\bf 1}$ & $0$ &$1$ \\
\hline
\hline
$S_3$ &  $\overline{\bf 3}$ & ${\bf 3}$ & $1/3$ & $-2$\\
\hline
\end{tabular}
\caption{ \footnotesize The model consists of three chiral fermion families, one vector-like fermion family and
two Higgs scalar doublets. The gauged $U(1)'$ is broken by a singlet $\phi$ leading to a massive $Z'$. 
We also include a scalar leptoquark triplet $S_3$.}
\label{model}
}
\end{minipage}\hfill
\begin{minipage}[b]{0.4\linewidth}
\centering
\includegraphics[width=30mm]{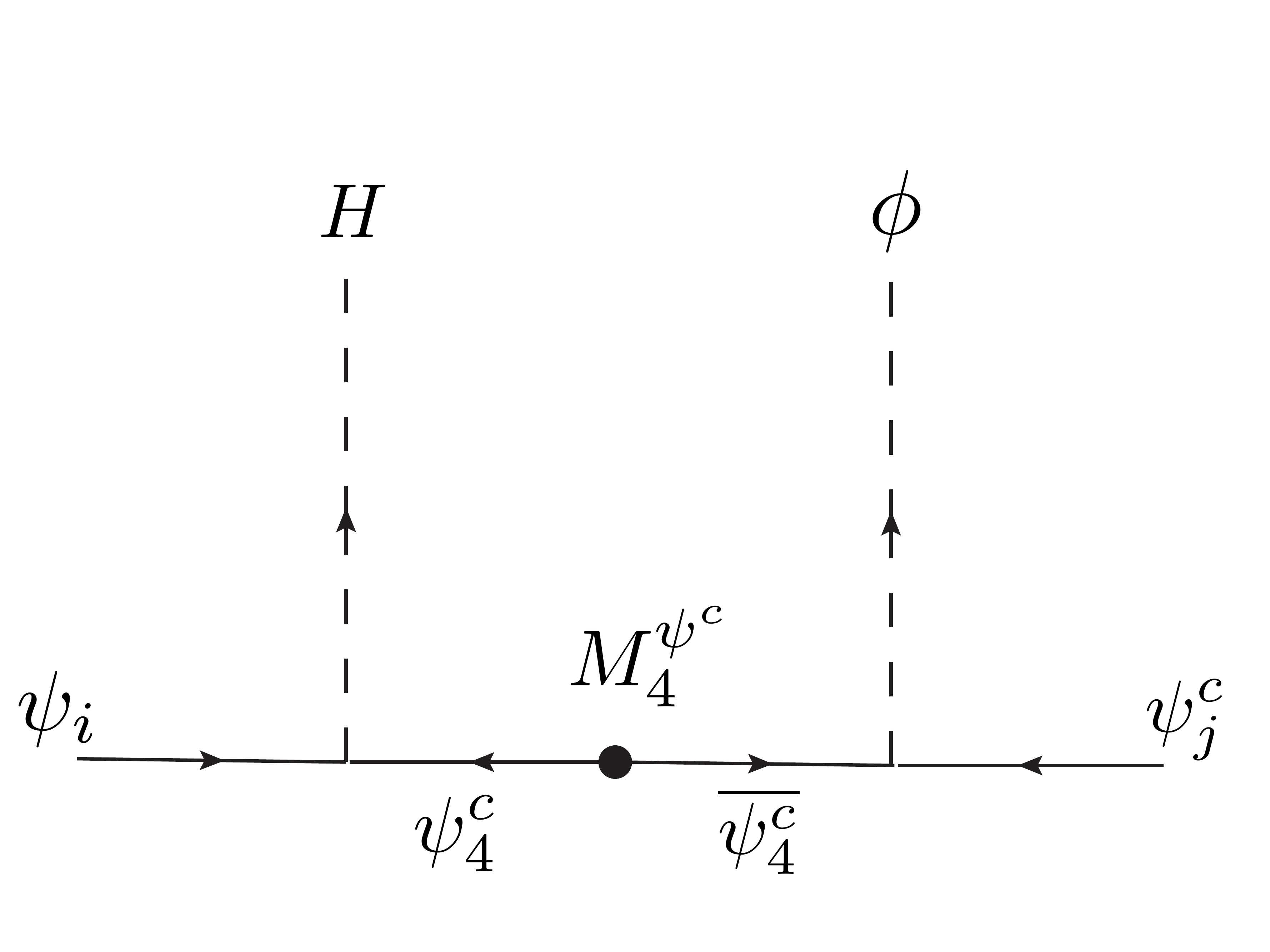}
\includegraphics[width=30mm]{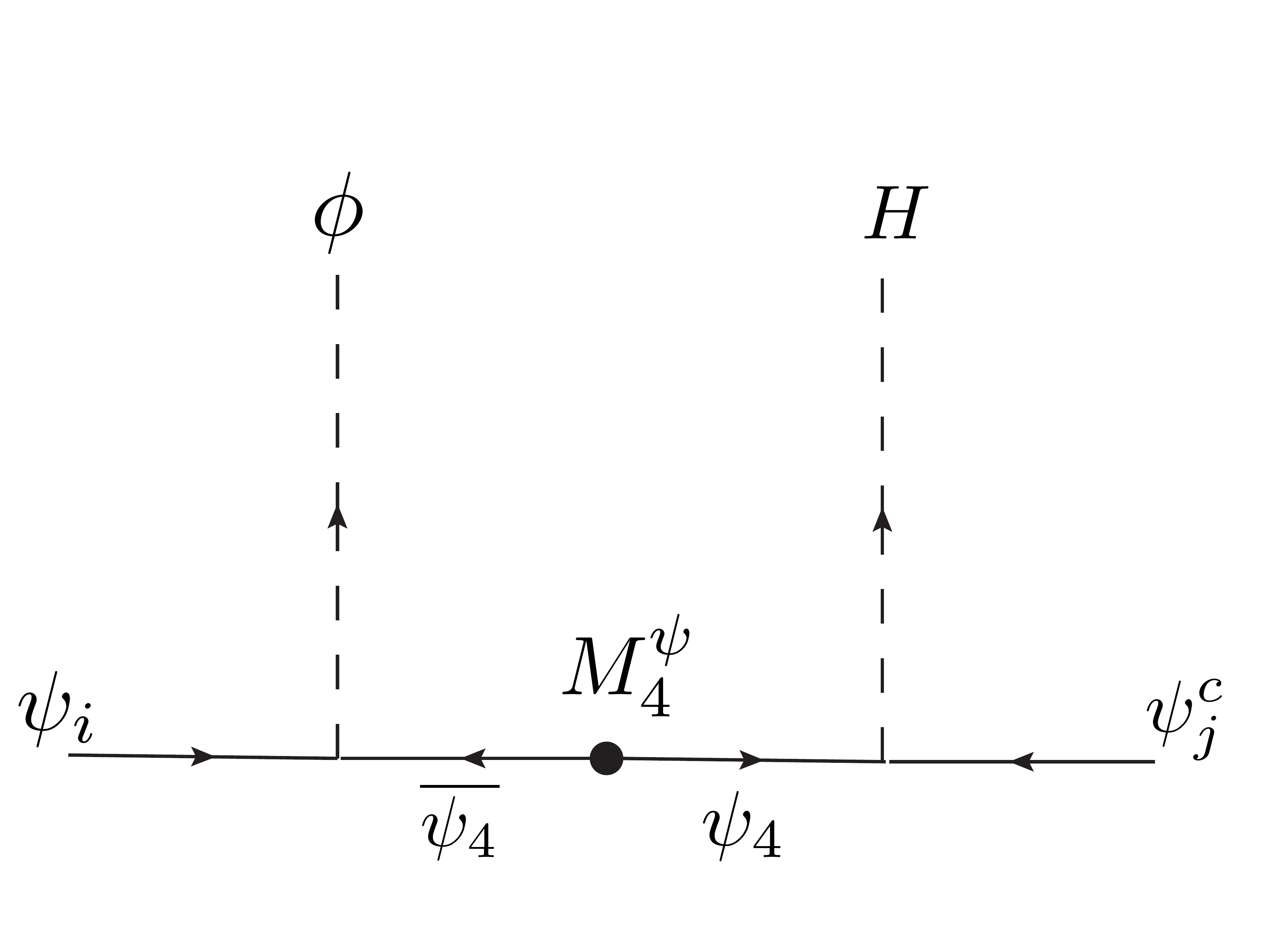}
\includegraphics[width=40mm]{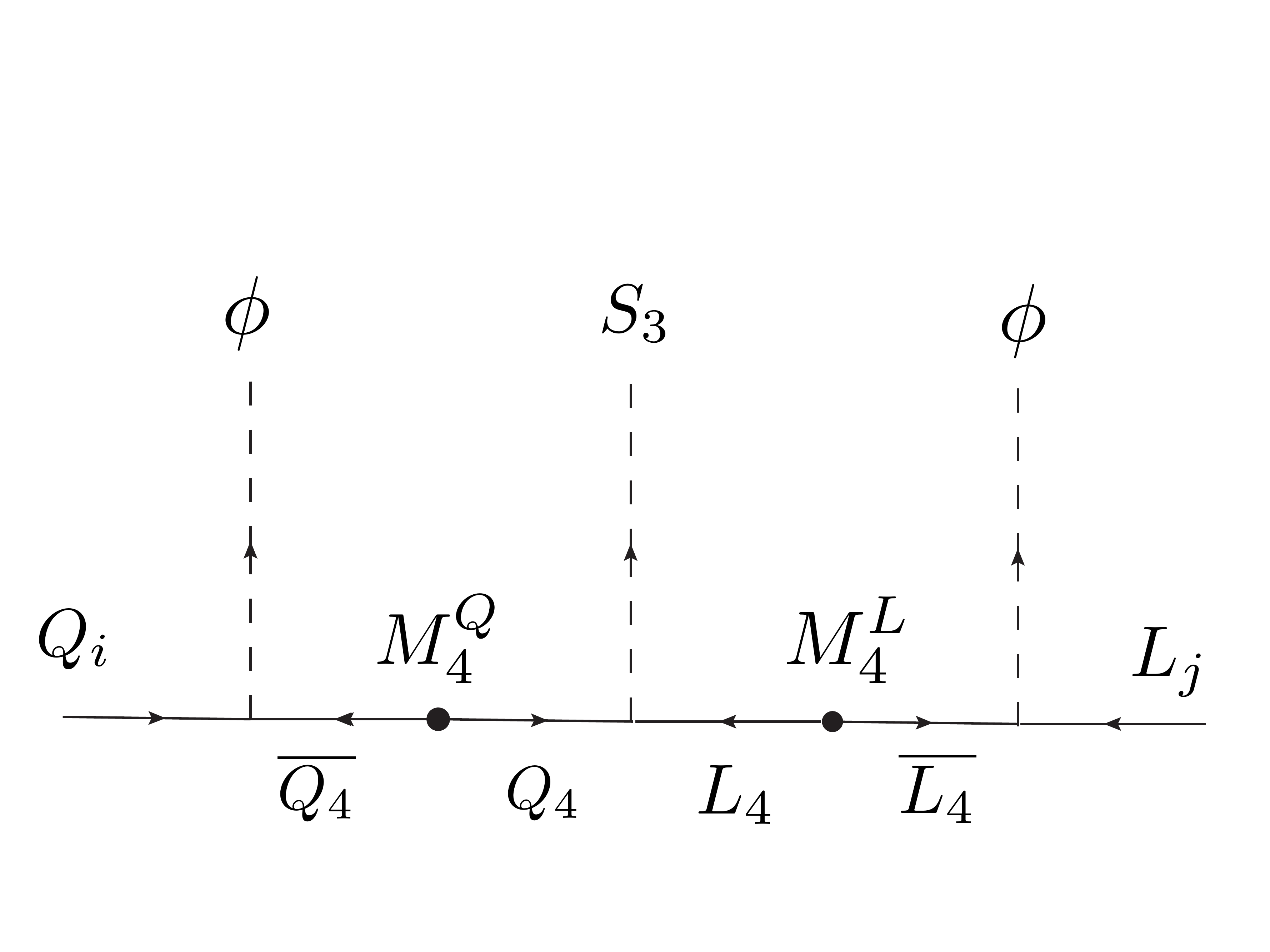}
\includegraphics[width=30mm]{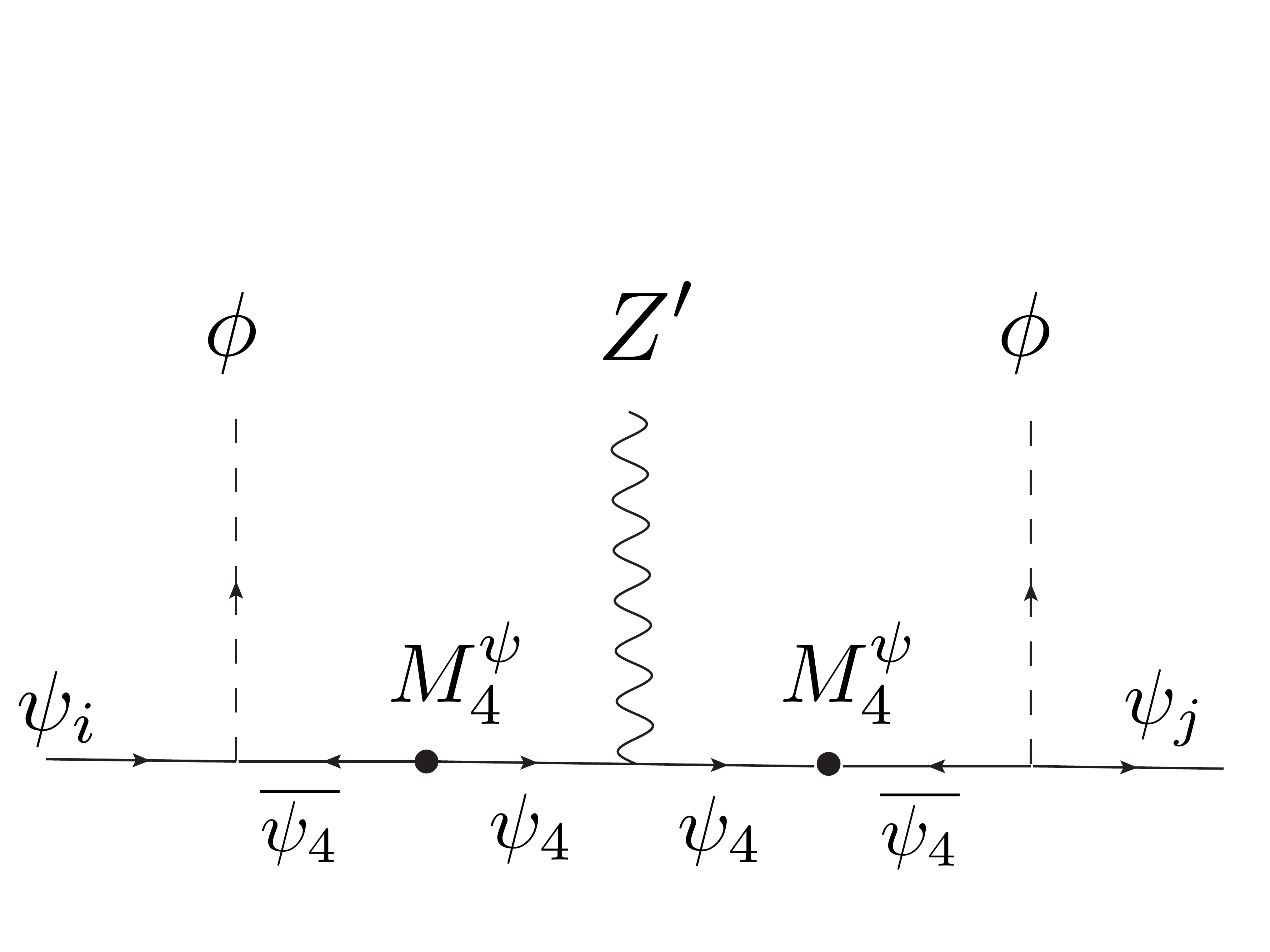}
\includegraphics[width=30mm]{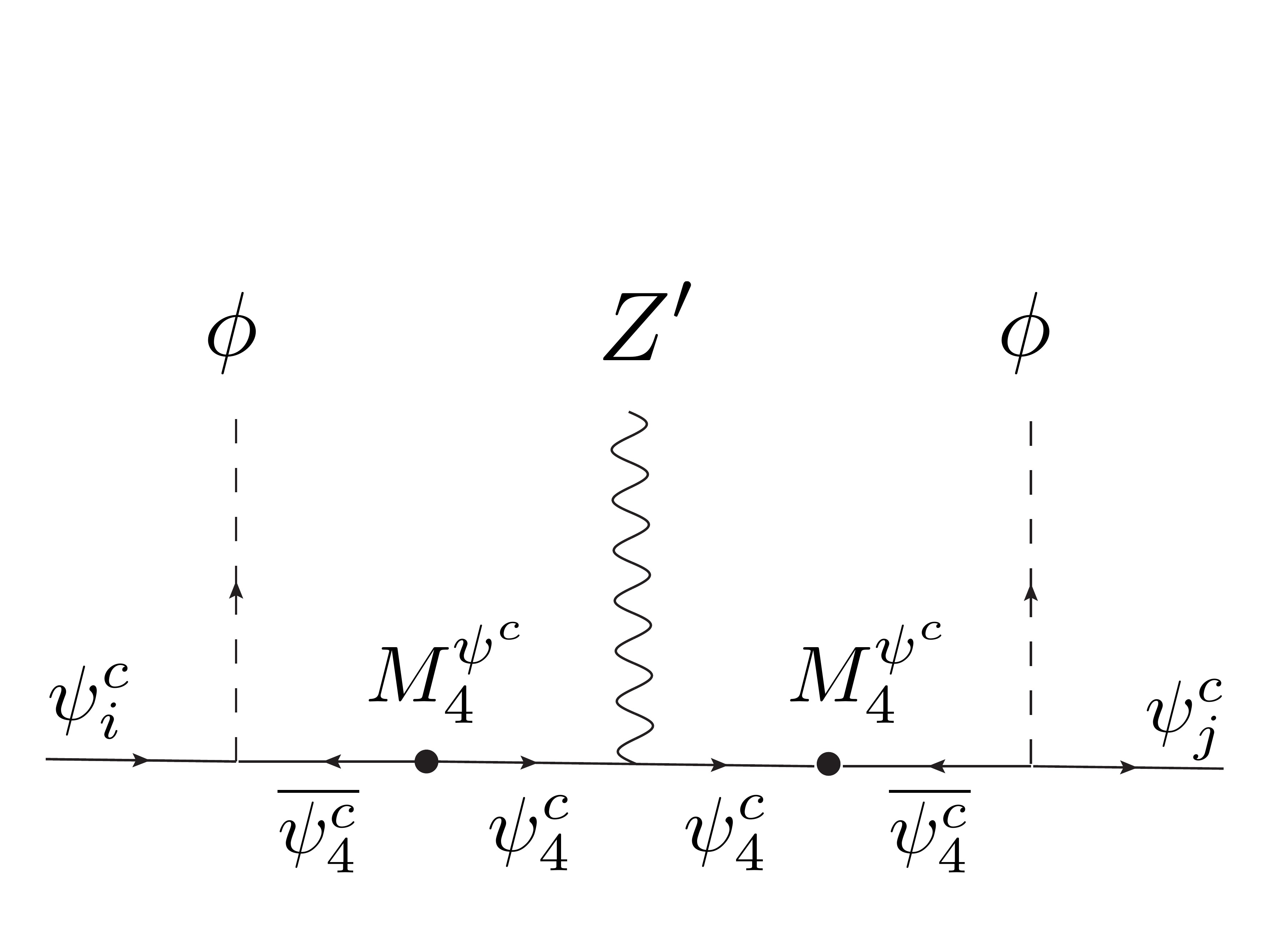}
  \caption{\footnotesize Effective Higgs couplings (upper)
(where $H=H_{u,d}$, $\psi_i=Q_i,L_i$ and 
$\psi^c_i =u^c_i,d^c_i,e^c_i$), leptoquark  couplings (middle) and $Z'$ couplings (lower).} \label{diagrams}
\end{minipage}
\end{table}

\subsection{The Higgs couplings}
We first consider the couplings involving the two Higgs doublets $H_{u,d}$. This was
first discussed in~\cite{Ferretti:2006df},
where a $Z_2$ symmetry prevented the usual Yukawa couplings. Here it is the gauged 
$U(1)'$ which forbids the usual Yukawa couplings since the Higgs carry the
new charges while the chiral fermions do not.
However Higgs scalar doublets with $U(1)'$ charge $-1$
can couple a chiral fermion to a vector-like fourth family fermion with $U(1)'$ charge $+1$,
controlled by new Yukawa couplings $y_{4i}$. The $U(1)'$ also allows the scalar singlet $\phi$ to couple 
a chiral fermion to a vector-like fourth family fermion,
controlled by new Yukawa couplings $x_{i}$. These couplings generate the 
$3\times3$ effective Yukawa matrices, via the upper diagrams in Table~\ref{diagrams},
in a particular basis~\cite{Ferretti:2006df}:
\begin{eqnarray}
 \label{eq:matrixU}
\!\!\!\!\!\!\!\!\!\!\!\!\!\!\!\!\!
y^{e,u}_{ij}=\left(\begin{array}{ccc}
0 & 0 & 0 \\
0 & \varepsilon^{e,u}_{22} &  \varepsilon^{e,u}_{23}\\
0 &  \varepsilon^{e,u}_{32} & y^{e,u}_{33}+\varepsilon^{e,u}_{33}
    \end{array}\right),
    \ \ \ \ 
    y^{d}_{ij}=\left(\begin{array}{ccc}
0 &  \varepsilon^{d}_{12} &  \varepsilon^{d}_{13} \\
0 &  \varepsilon^{d}_{22} &  \varepsilon^{d}_{23}\\
0 &  \varepsilon^{d}_{32} & y^{d}_{33}+ \varepsilon^{d}_{33}
    \end{array}\right),\ \ \ \ 
    \label{Yuk_effective}
    \end{eqnarray}   
  where the effective Yukawa couplings $ \varepsilon_{ij}$ are defined as $\varepsilon^e_{ij} H_dL_ie^c_j$,
$\varepsilon_{ij}^{u}H_uQ_iu^c_j$, $\varepsilon_{ij}^{d}H_dQ_id^c_j$,
and are given by the upper left diagrams in Table~\ref{diagrams},
hence $\varepsilon^{e,u,d}_{ij}\propto 1/M^{e^c,u^c,d^c}_{4}$.
These couplings are suppressed $ \varepsilon_{ij} \ll y_{33}$,
assuming $M^{L,Q}_{4}\ll M^{e^c,u^c,d^c}_{4}$ (see~\cite{Ferretti:2006df} for more details).

  To leading order the dominant third family Yukawa couplings are given by
  the upper right diagrams in Table~\ref{diagrams},
   \begin{equation}
y_{\tau} \approx y^{e}_{33}\approx y^{e}_{43} 
\left(\frac{x^{L}_{3} \langle \phi \rangle }{M^{L}_{4}}\right), \ \ \ \ 
y_t\approx y^{u}_{33}\approx y^{u}_{43} 
\left(\frac{x^{Q}_{3}  \langle \phi \rangle }{M^{Q}_{4}}\right), \ \ \ \ 
y_b\approx y^{d}_{33}\approx y^{d}_{43} 
\left( \frac{x^{Q}_{3} \langle \phi \rangle }{M^{Q}_{4}}\right)
\label{33}
\end{equation}
where the effective Yukawa couplings are defined for the two Higgs doublet model as $y^{e}_{33}H_dL_3e^c_3$,
$y^{u}_{33}H_uQ_3u^c_3$ and $y^{d}_{33}H_dQ_3d^c_3$. In this basis, only the third family Yukawa couplings
originate from such diagrams~\cite{Ferretti:2006df}.

\begin{figure}
\begin{minipage}{0.33\linewidth}
\centerline{\includegraphics[width=0.95\linewidth=true]{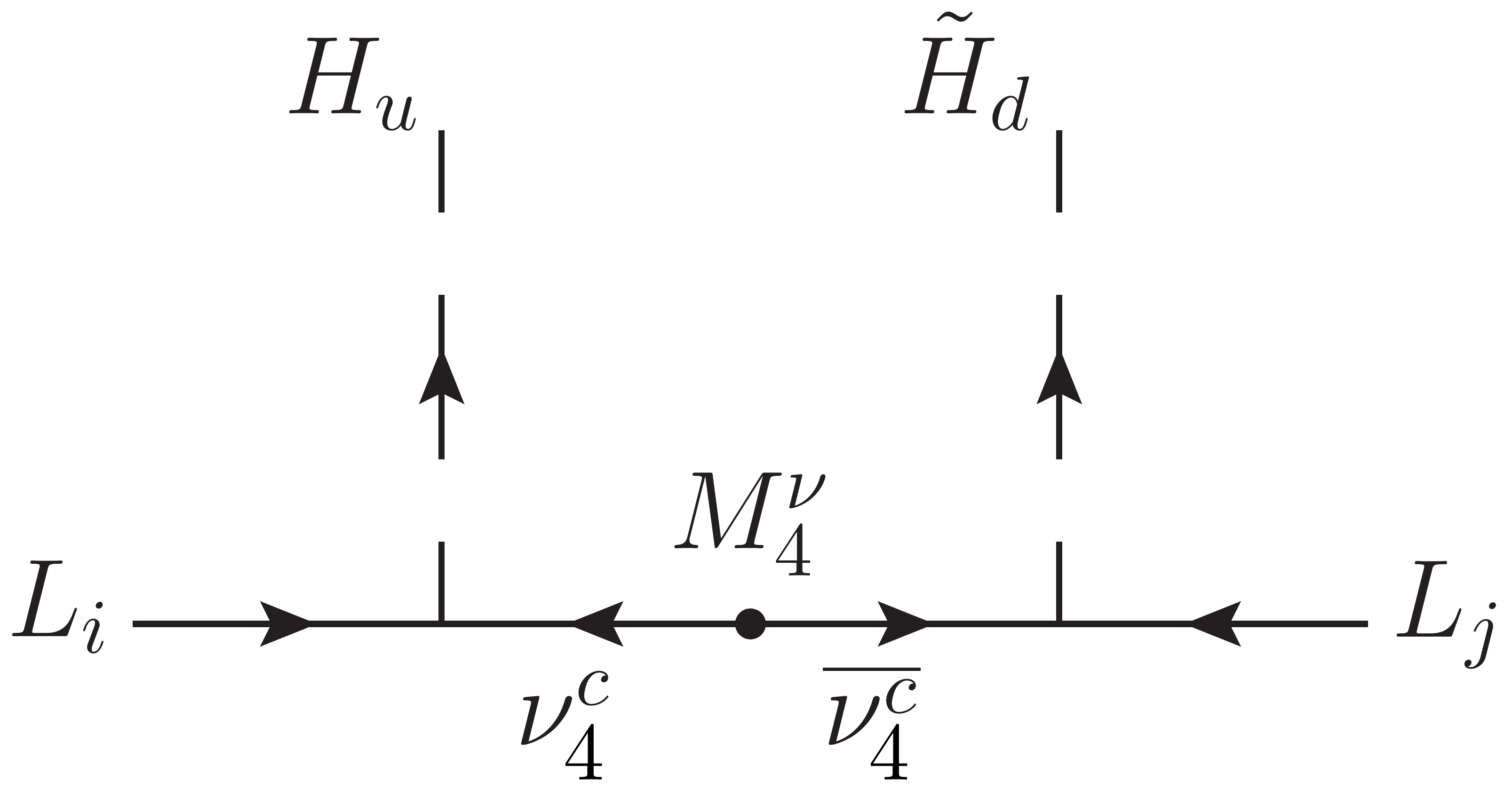}}
\end{minipage}
\begin{minipage}{0.32\linewidth}
\centerline{\includegraphics[width=0.8\linewidth]{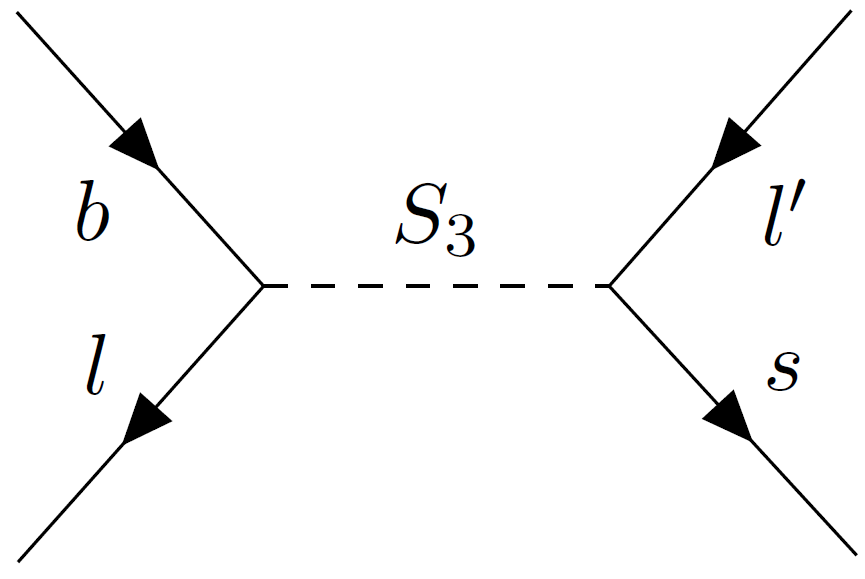}}
\end{minipage}
\begin{minipage}{0.33\linewidth}
\centerline{\includegraphics[width=0.8\linewidth=true]{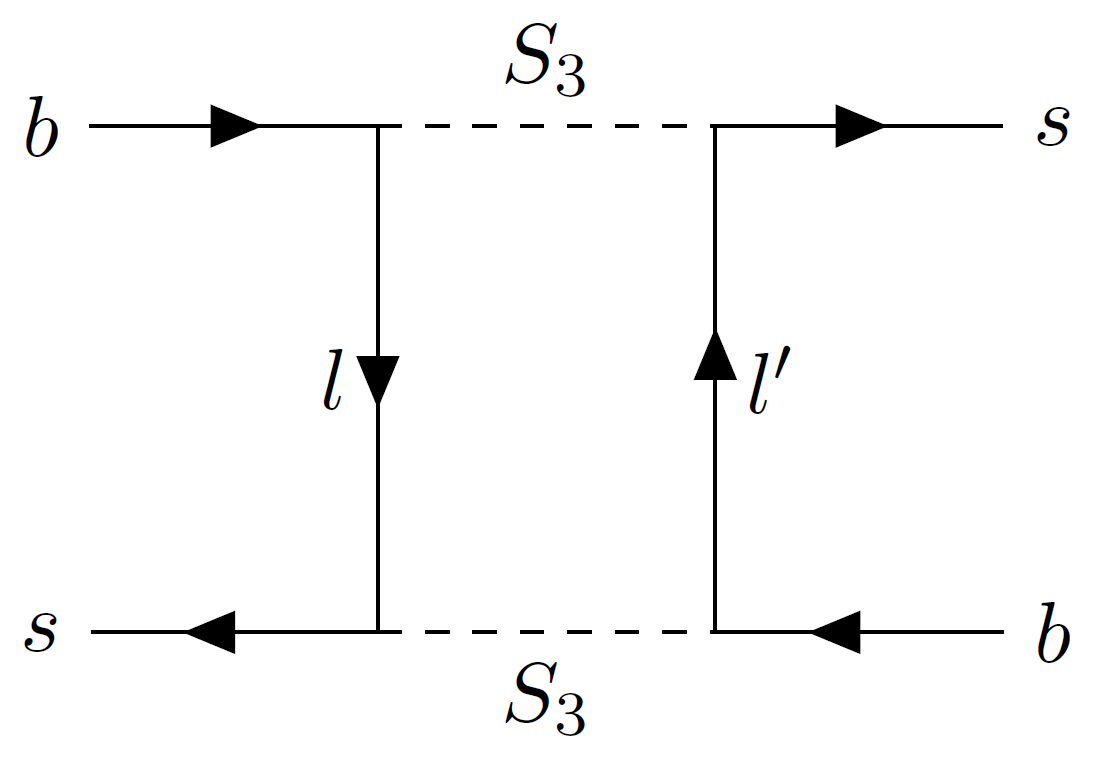}}
\end{minipage}
\caption{The fourth family vector-like neutrinos allows a new contribution to neutrino mass
via a diagram involving two different Higgs doublets 
$H_u$, $\tilde H_d$ (left), which we refer to as the type Ib seesaw mechanism.
The leptoquark $S_3$ contributes to $R_{K^{(^*)}}$ at tree-level (centre), and to $B_s$ mixing 
at one loop (right). }
\label{pheno1}
\end{figure}

Interestingly, the fourth family vector-like neutrinos provide a new contribution to neutrino mass
via the type Ib seesaw\footnote{We refer to the seesaw mechanism involving two different Higgs doublets 
$H_u$, $\tilde H_d$
as type Ib to distinguish it from the usual seesaw mechanism involving
two identical Higgs doublets $H_u$ which we refer to as type Ia.} diagram in Fig.~\ref{pheno1} (left)~\cite{Hernandez-Garcia:2019uof}.
Below the mass scale of the fourth family of vector-like neutrinos,
this leads to a new Weinberg operator for neutrino mass of the form $\frac{1}{M^{\nu_c}_4}H_u\tilde H_d L_iL_j$ involving the two 
different Higgs doublets $H_u$, $\tilde H_d$, where 
the charge conjugated doublet $\tilde H_d=-i\sigma_2 H_d^*$, and $H_d^*$ is the complex conjugate of $H_d$.
For more details including a phenomenological analysis see~\cite{Hernandez-Garcia:2019uof}.

\subsection{The leptoquark couplings }
We now consider the couplings involving the scalar leptoquark triplet $S_3$
as discussed in~\cite{deMedeirosVarzielas:2019okf}.
The assigned $U(1)'$ charges allow the renormalisable leptoquark coupling, $\lambda_4 S_3Q_4L_4$,
involving the fourth family, but not the first three families. The
middle diagram in Table~\ref{diagrams} generates 
a single effective leptoquark coupling, which involves the third family (only) in the same
basis as Eq.\ref{Yuk_effective}~\cite{deMedeirosVarzielas:2019okf}:
\be
\lambda_4 
\left(\frac{x^{L}_{3} \langle \phi \rangle }{M^{L}_{4}}\right)
\left(\frac{x^{Q}_{3}  \langle \phi \rangle }{M^{Q}_{4}}\right)
S_3Q_3L_3
\approx
\lambda_4 
\left(\frac{y^{e}_{33}}{y^{e}_{43} }\right)
\left(\frac{y^{u}_{33}}{y^{u}_{43} }\right)
S_3Q_3L_3
\approx y_{\tau} y_tS_3Q_3L_3
\label{leptoquark_couplings_33}
\ee
where the first equality in Eq.\ref{leptoquark_couplings_33} has used Eq.\ref{33}, and the second equality sets 
$y_{4i}\approx \lambda_4 \approx 1$.

Effective leptoquark couplings to first and second family quarks and leptons are generated
when the Yukawa matrices in Eq.\ref{Yuk_effective}
are diagonalised and so are suppressed by $ \varepsilon_{ij} / y_{33}$.
Since down quark mixing is larger than up quark mixing (due to the milder mass hierarchy), we assume $\theta_{23}^d\approx V_{ts} $, while the analogous charged lepton mixing angle $\theta_{23}^e$ is similarly small.
Hence in the diagonal Yukawa basis we have leptoquark couplings involving the
left-handed lepton doublets $L_3=(\nu_{\tau}, \tau )^T_L$, $L_2=(\nu_{\mu}, \mu )^T_L$, and quark doublets $Q_3=(t, b)^T_L$, $Q_2=(c, s)^T_L$, from Eq.\ref{leptoquark_couplings_33}, assuming $y_t\approx 1$,
\be
y_{\tau} S_3Q_3L_3, \ \ \ \ y_{\tau} V_{ts} S_3Q_2L_3, \ \ \ \ 
y_{\tau} \theta_{23}^e S_3Q_3L_2,   \ \ \ \ y_{\tau} \theta_{23}^e  V_{ts} S_3Q_2L_2, \ \ \ \ \ldots
\label{leptoquark_couplings_ij}
\ee
Thus, after a number of reasonable dynamical assumptions, we have obtained the leptoquark couplings 
in Eq.\ref{leptoquark_couplings_ij} in terms of  Yukawa couplings and mixing angles.

The leptoquark couplings in Eq.\ref{leptoquark_couplings_ij} have a number of interesting phenomenological implications,
mainly due to the the couplings of 
the electric charge $+4/3$ component of $S_3$ to the physical 
left-handed down quark and charged lepton mass
eigenstates
\be
\lambda_{b\tau }  S_3 b_L \tau_L , \ \ \ \  \lambda_{s\tau } S_3 s_L \tau_L , \ \ \ \ 
\lambda_{b\mu }   S_3b_L \mu_L ,   \ \ \ \ 
\lambda_{s\mu } S_3 s_L \mu_L , \ \ \ \ 
\label{leptoquark_couplings_deij}
\ee
\be
\lambda_{b\tau } \approx y_{\tau}, \ \ \ \  \lambda_{s\tau } \approx y_{\tau} V_{ts},\ \ \ \ 
\lambda_{b\mu }  \approx y_{\tau} \theta_{23}^e, \ \ \ \ \lambda_{s\mu }  \approx y_{\tau} \theta_{23}^e  V_{ts}.
\label{leptoquark_couplings_lambda}
\ee

The leptoquark $S_3$ contributes to $R_{K^{(^*)}}$ at tree-level, 
via the (centre) diagram in Fig.\ref{pheno1},
where the requirement to explain the anomaly is~\cite{deMedeirosVarzielas:2019okf}  
\be
\label{RKuirement}
\frac{\lambda_{b\mu} \lambda_{s\mu} }{M_{S_3}^2}\approx \frac{y_{\tau}^2  (\theta_{23}^e)^2 V_{ts}}{M_{S_3}^2} \approx
\frac{1.1}{(35~\rm{TeV})^2} 
\rightarrow 
y_{\tau}^2  (\theta_{23}^e)^2 \approx
2.2\times 10^{-2}  \left( \frac{M_{S_3}}{1~\rm{TeV}} \right)^2,
\ee
using $V_{ts} \approx 4.0\times 10^{-2}$, which requires quite a large $y_{\tau}\approx 1$ 
(i.e. large $\tan \beta =\langle H_u \rangle / \langle H_d \rangle  $) and a large mixing angle $\theta_{23}^e \approx 0.1$,
together with a low leptoquark mass $M_{S_3}\approx 1$ TeV, close to current LHC limits~\cite{deMedeirosVarzielas:2019okf}.

On the other hand, $B_s$ mixing only occurs at one loop, via the (right) diagram in Fig.\ref{pheno1},
dominated by $\tau $ exchange,
leading to the 2015 bound~\cite{deMedeirosVarzielas:2019okf}
\be
\frac{ (\lambda_{s\tau}\lambda_{b \tau})^2}{16\pi^2 M^2_{S_3}}
 \approx 
 \frac{y^4_{\tau} V_{ts}^2  }{16\pi^2 M^2_{S_3}}
 \leq \frac{1}{(140~\rm{TeV})^2}
  \rightarrow 
 y^4_{\tau} \leq 
 5.0 \left( \frac{M_{S_3}}{1~\rm{TeV}} \right)^2 
\ee
which is satisfied even for large $y_{\tau}\approx 1$ with $M_{S_3}\approx 1$ TeV. 
However the stronger 2017 bound with scale of $770$ TeV instead of $140$ TeV implies a bound of $y^4_{\tau} \leq 0.16$
for $M_{S_3}\approx 1$ TeV~\cite{deMedeirosVarzielas:2019okf}.

\subsection{The $Z'$ couplings}
\begin{figure}
\begin{minipage}{0.33\linewidth}
\centerline{\includegraphics[width=0.9\linewidth]{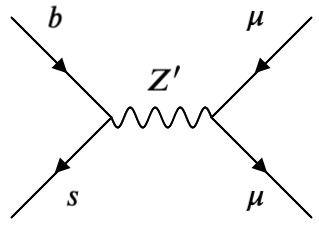}}
\end{minipage}
\begin{minipage}{0.33\linewidth}
\centerline{\includegraphics[width=0.9\linewidth=true]{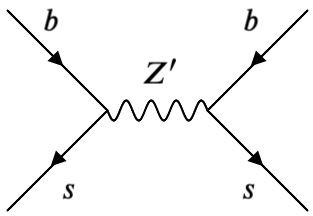}}
\end{minipage}
\begin{minipage}{0.33\linewidth}
\centerline{\includegraphics[width=0.9\linewidth]{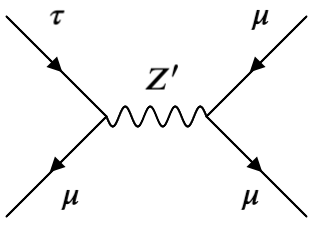}}
\end{minipage}
\caption{These $Z'$ exchange diagrams 
contribute to $R_{K^{(^*)}}$ (left), to $B_s$ mixing (centre)
and to $\tau\rightarrow \mu \mu \mu $ (right).
The couplings are defined as $g_{bs}Z'_{\mu}b_L^{\dagger}\gamma^{\mu}s_L$,
$g_{\mu \mu}Z'_{\mu}\mu_L^{\dagger}\gamma^{\mu}\mu_L$ and $g_{\tau \mu}Z'_{\mu}\tau_L^{\dagger}\gamma^{\mu}\mu_L$.}
\label{pheno2}
\end{figure}

We now consider the couplings involving the $Z'$ as discussed in~\cite{King:2018fcg}.
Although the chiral fermions do not carry $U(1)'$ charges, the lower diagrams in Table~\ref{diagrams} generate effective $Z'$ couplings to chiral fermions, via the vector-like fourth family fermions which do carry $U(1)'$ charges
(which are trivially anomaly free).
The $Z'$ couplings in the basis of Eq.\ref{Yuk_effective} are dominated by left-handed couplings
to the third family~\cite{King:2018fcg},
\footnote{
Unlike the leptoquark case, there are also small $Z'$ couplings to the right-handed fermions,
in the basis of Eq.\ref{Yuk_effective},
$ \varepsilon^{u}_{ij} \varepsilon^{u}_{ji}  g' Z'_{\mu}u^{c \dagger}_i\gamma^{\mu}u^c_j$,
$\varepsilon^{d}_{ij} \varepsilon^{d}_{ji} g' Z'_{\mu}d^{c \dagger}_i\gamma^{\mu}d^c_j$,
$\varepsilon^{e}_{ij} \varepsilon^{e}_{ji} g' Z'_{\mu}e^{c \dagger}_i\gamma^{\mu}e^c_j$,
which are second order in the small Yukawa couplings. 
Thus, the small value of the Yukawa coupling of the charm quark 
implies that the $c_R$ coupling to $Z'$ is suppressed by $(m_c/m_t)^2\sim 10^{-4}$ in this basis.
The $s_R$ coupling to $Z'$ is similarly suppressed, so there there is a negligible contribution to $K_0-\bar{K_0}$ mixing
for $M_Z'\sim 1$ TeV.
}
\be
y_t^2 g' Z'_{\mu}Q^{\dagger}_3\gamma^{\mu}Q_3+
y_{\tau}^2 g' Z'_{\mu}L^{\dagger}_3\gamma^{\mu}L_3,
\label{Zp_Q3L3}
\ee
The $Z'$ couplings in Eq.\ref{Zp_Q3L3}
are analogous to the case of the third family leptoquark couplings in Eq.\ref{leptoquark_couplings_33},
with no couplings to the first or second family in the basis 
of Eq.\ref{Yuk_effective}. However, flavour changing couplings involving the quark doublets 
$Q_3=(t, b)^T_L$, $Q_2=(c, s)^T_L$, will be generated
when the Yukawa matrices in Eq.\ref{Yuk_effective} are diagonalised,
 \be
y_t^2 g' Z'_{\mu}Q^{\dagger}_3\gamma^{\mu}Q_3 \ \ \ \ 
\rightarrow \ \ \ \ 
V_{ts}Z'_{\mu}Q^{\dagger}_3\gamma^{\mu}Q_2,\ \ \ \ 
V_{ts}^2Z'_{\mu}Q^{\dagger}_2\gamma^{\mu}Q_2,\ \ \ \ \ldots \ \ \ \ 
\rightarrow \ \ \ \ 
V_{ts}Z'_{\mu}b_L^{\dagger}\gamma^{\mu}s_L, \ \ \ \ \ldots
\label{Zp_Q_couplings_ij}
\ee
Similarly the operator $y_{\tau}^2 g' Z'_{\mu}L^{\dagger}_3\gamma^{\mu}L_3$ 
in Eq.\ref{Zp_Q3L3} leads to 
flavour changing couplings involving the
lepton doublets $L_3=(\nu_{\tau}, \tau )^T_L$, $L_2=(\nu_{\mu}, \mu )^T_L$,
controlled by a left-handed lepton mixing $\theta_{23}^e$,
\be
\! \! \! \! 
\theta_{23}^ey_{\tau}^2 Z'_{\mu}L^{\dagger}_3\gamma^{\mu}L_2,\ \ 
(\theta_{23}^e)^2y_{\tau}^2 Z'_{\mu}L^{\dagger}_2\gamma^{\mu}L_2 
\ \ \ \ \ldots \ \ \ \ 
\rightarrow \ \ \ \ 
\theta_{23}^ey_{\tau}^2 Z'_{\mu}\tau_L^{\dagger}\gamma^{\mu}\mu_L,\ \ 
(\theta_{23}^e)^2y_{\tau}^2 Z'_{\mu}\mu_L^{\dagger}\gamma^{\mu}\mu_L  
\label{Zp_L_couplings_ij}
\ee
where we have taken $y_t\approx g'\approx 1$. 
The couplings in Eqs.\ref{Zp_Q_couplings_ij}, \ref{Zp_L_couplings_ij}
control the $Z'$ exchange diagrams in Fig.\ref{pheno2} which 
contribute to $R_{K^{(^*)}}$ (left), to $B_s$ mixing (centre)
and to $\tau\rightarrow \mu \mu \mu $ (right).

The $Z'$ contributes to $R_{K^{(^*)}}$ at tree-level, 
via the (left) diagram in Fig.\ref{pheno2},
where the requirement to explain the anomaly (ignoring the contribution from the leptoquark)
 is~\cite{King:2018fcg}
\be
\label{RKuirement2}
\frac{g_{\mu \mu} g_{bs} }{M_{Z'}^2}\approx \frac{y_{\tau}^2  (\theta_{23}^e)^2 V_{ts}}{M_{Z'}^2} \approx
\frac{1.1}{(35~\rm{TeV})^2} 
\rightarrow 
y_{\tau}^2  (\theta_{23}^e)^2 \approx
2.2\times 10^{-2}  \left( \frac{M_{Z'}}{1~\rm{TeV}} \right)^2,
\ee
using $V_{ts} \approx 4.0\times 10^{-2}$, which is analogous to the expression we obtained for the leptoquark
in Eq.\ref{RKuirement}.
As before, this requires quite a large $y_{\tau}\approx 1$ 
(i.e. large $\tan \beta =\langle H_u \rangle / \langle H_d \rangle  $) and a large mixing angle $\theta_{23}^e \approx 0.1$,
together with a low mass $M_{Z'}\approx 1$ TeV, close to current LHC limits~\cite{Falkowski:2018dsl}.


Now $B_s$ mixing is mediated by tree-level $Z'$ exchange as in the (centre) diagram in Fig.\ref{pheno2},
leading to the 2015 bound~\cite{Falkowski:2018dsl},
\be
\frac{g_{bs} g_{bs} }{M_{Z'}^2}
 \approx 
  \frac{ V_{ts}^2  }{M_{Z'}^2}
 \leq \frac{1}{(140~\rm{TeV})^2}
  \rightarrow 
M_{Z'}\geq V_{ts}(140~\rm{TeV})= 5.6~\rm{TeV}
\ee
However the stronger 2017 bound with scale of $770$ TeV instead of $140$ TeV implies a bound of 
$M_{Z'}\geq 31~\rm{TeV}$, which seems incompatible with the $R_{K^{(^*)}}$ requirement in Eq.\ref{RKuirement2}.

Moreover $\tau\rightarrow \mu \mu \mu $ is mediated by tree-level $Z'$ exchange as in the (right) diagram in Fig.\ref{pheno2},
leading to the bound~\cite{Falkowski:2018dsl},
\be
\frac{g_{\tau \mu} g_{\mu \mu} }{M_{Z'}^2}
 \approx 
  \frac{(\theta_{23}^e)^3y_{\tau}^4 }{M_{Z'}^2}
 \leq \frac{1}{(16~\rm{TeV})^2}
 \rightarrow 
y_{\tau}^4  (\theta_{23}^e)^3 \leq
4.0\times 10^{-3}  \left( \frac{M_{Z'}}{1~\rm{TeV}} \right)^2.
\ee

Writing $g_{\tau \mu} = g_{\mu \mu}/\theta^e_{23}$,
the bounds on $B_s$ mixing and $\tau\rightarrow \mu \mu \mu $ may be written as:
\be
\frac{g_{bs}}{M_{Z'}}
 \leq \frac{1}{(140~\rm{TeV})}, \ \ \ \ 
 \frac{g_{\mu \mu} }{M_{Z'}}
  \leq \frac{(\theta^e_{23})^{1/2}}{(16~\rm{TeV})}
\ee
which may be combined, leading to a bound~\footnote{I am grateful to E.Perdomo for pointing out this bound.} on the contribution to $R_{K^{(^*)}}$,
\be
\frac{g_{\mu \mu} }{M_{Z'}} \frac{g_{bs}}{M_{Z'}}
 \leq \frac{(\theta^e_{23})^{1/2}}{(140~\rm{TeV})(16~\rm{TeV})}
 =\frac{(\theta^e_{23})^{1/2}}{(47~\rm{TeV})^2}
 \label{combined_bound}
 \ee
 which is somewhat less than the $\frac{1.1}{(35~\rm{TeV})^2}$ 
 required in Eq.\ref{RKuirement2} to explain the anomaly.
Moreover, the stronger 2017 bound with scale of $770$ TeV instead of $140$ TeV implies a bound of 
$\frac{(\theta^e_{23})^{1/2}}{(111~\rm{TeV})^2}$, which is significantly less than
the $\frac{1.1}{(35~\rm{TeV})^2}$ required to explain the anomaly.

\section{Summary and Conclusion}
\label{conclusion}

In this talk we have explored the possibility that Higgs Yukawa couplings are related to the 
couplings of a new scalar triplet leptoquark or $Z'$, providing a predictive theory of flavour, including flavour changing,
and flavour non-universality. 

In particular, we have here combined (for the first time) the $Z'$ model~\cite{King:2018fcg}
and the leptoquark model~\cite{deMedeirosVarzielas:2019okf}, including also a
vector-like fourth family of fermions, as a possible explanation of $R_{K^{(^*)}}$.
The idea of these models is that the Yukawa couplings are generated by the same physics that generates the
$Z'$ and leptoquark couplings, namely mixing with the vector-like fourth family.
The combined model proposed here allows
 $Z'$ and leptoquark contributions to be compared in the same framework.

In the combined model considered here, the leptoquark couplings in Eq.\ref{leptoquark_couplings_deij} are given in terms of
Yukawa couplings and mixing angles in Eq.\ref{leptoquark_couplings_lambda}.
We have seen that such a flavoured leptoquark can just about account for $R_{K^{(^*)}}$,
while satisfying the bound on $B_s$ mixing, providing that the leptoquark mass is close to its
current LHC bound, namely $M_{S_3}\approx 1$ TeV, which is a clear prediction of the model,
providing a target future LHC runs. However, as seen above, this also requires quite a large $y_{\tau}\approx 1$ 
and a large mixing angle $\theta_{23}^e \approx 0.1$, so the leptoquark-only explanation of $R_{K^{(^*)}}$
 is under some tension.

Unfortunately, the $Z'$ couplings in Eqs.~\ref{Zp_Q_couplings_ij},\ref{Zp_L_couplings_ij}, 
are unable to account for $R_{K^{(^*)}}$, while satisfying the bounds from $B_s$ mixing and 
$\tau\rightarrow \mu \mu \mu $, as can be seen from the bound in Eq.\ref{combined_bound}.
However it is possible that other (lepton) Yukawa matrix structures may allow $Z'$ exchange to account for $R_{K^{(^*)}}$,
as previously discussed~\cite{King:2018fcg}.

There is an important distinction to be drawn between the leptoquark and $Z'$ explanations of $R_{K^{(*)}}$,
when linked to a theory of flavour.
The leptoquark mass term $M_{S_3}^2|S_3|^2$ respects
$U(1)'$, and hence the leptoquark mass $M_{S_3}$ is independent of the order parameter
$\langle \phi \rangle$.
As far as the leptoquark mass is concerned, the flavour breaking scale 
$\langle \phi \rangle$
may be anywhere from the Planck scale to the electroweak scale,
providing that the ratios $\langle \phi \rangle /M_4$ which govern the Yukawa couplings are held fixed.
By contrast, since the $Z'$ mass is of order $\langle \phi \rangle$, a TeV scale $Z'$ (as would be required to 
account for $R_{K^{(*)}}$) would require a theory of flavour near the electroweak scale!

\section*{Acknowledgments}
I am grateful to the organisers for arranging such a stimulating conference.
Work supported by the STFC Consolidated Grant ST/L000296/1 and the European Union's Horizon 2020 Research and Innovation programme under Marie Sk\l{}odowska-Curie grant agreements 
Elusives ITN No.\ 674896 and InvisiblesPlus RISE No.\ 690575.

\end{document}